\documentclass[8.5pt,twoside,twocolumn]{article}
\usepackage[super,sort&compress,comma]{natbib} 
\usepackage{mhchem}
\usepackage{times,mathptmx}
\usepackage{sectsty}
\usepackage{balance} 
\usepackage{ulem}

\usepackage{graphicx} %eps figures can be used instead
\usepackage{lastpage}
\usepackage[format=plain,justification=raggedright,singlelinecheck=false,font=small,labelfont=bf,labelsep=space]{caption} 
\usepackage{fancyhdr}
\usepackage{color}

\begin{document}

\thispagestyle{plain}
\fancypagestyle{plain}{
\fancyhead[L]{\includegraphics[height=8pt]{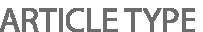}}
\fancyhead[C]{\hspace{-1cm}\includegraphics[height=20pt]{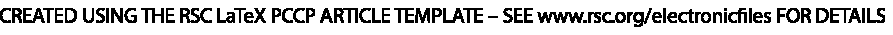}}
\fancyhead[R]{\includegraphics[height=10pt]{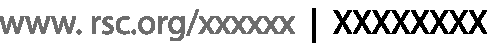}\vspace{-0.2cm}}
\renewcommand{\headrulewidth}{1pt}}
\renewcommand{\thefootnote}{\fnsymbol{footnote}}
\renewcommand\footnoterule{\vspace*{1pt}% 
\hrule width 3.4in height 0.4pt \vspace*{5pt}} 
\setcounter{secnumdepth}{5}

\makeatletter 
\def\subsubsection{\@startsection{subsubsection}{3}{10pt}{-1.25ex plus -1ex minus -.1ex}{0ex plus 0ex}{\normalsize\bf}} 
\def\paragraph{\@startsection{paragraph}{4}{10pt}{-1.25ex plus -1ex minus -.1ex}{0ex plus 0ex}{\normalsize\textit}} 
\renewcommand\@biblabel[1]{#1}            
\renewcommand\@makefntext[1]% 
{\noindent\makebox[0pt][r]{\@thefnmark\,}#1}
\makeatother 
\renewcommand{\figurename}{\small{Fig.}~}
\sectionfont{\large}
\subsectionfont{\normalsize} 

\fancyfoot{}
\fancyfoot[LO,RE]{\vspace{-7pt}\includegraphics[height=9pt]{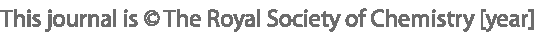}}
\fancyfoot[CO]{\vspace{-7.2pt}\hspace{12.2cm}\includegraphics{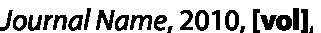}}
\fancyfoot[CE]{\vspace{-7.5pt}\hspace{-13.5cm}\includegraphics{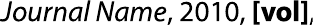}}
\fancyfoot[RO]{\footnotesize{\sffamily{1--\pageref{LastPage} ~\textbar  \hspace{2pt}\thepage}}}
\fancyfoot[LE]{\footnotesize{\sffamily{\thepage~\textbar\hspace{3.45cm} 1--\pageref{LastPage}}}}
\fancyhead{}
\renewcommand{\headrulewidth}{1pt} 
\renewcommand{\footrulewidth}{1pt}
\setlength{\arrayrulewidth}{1pt}
\setlength{\columnsep}{6.5mm}
\setlength\bibsep{1pt}

\twocolumn[
  \begin{@twocolumnfalse}
\noindent\LARGE{\textbf{Energy barriers and cell migration in densely packed tissues$^\dag$}}
\vspace{0.6cm}
%\textit{$^{a}$} \textit{$^{b\ddag}$}

\noindent\large{\textbf{Dapeng Bi$^{\ast}$, J. H. Lopez$^{\ast}$,  J. M. Schwarz$^{\ast}$ and
M. Lisa Manning$^{\ast}$\textit{$^{a}$}}}\vspace{0.5cm}
%Please note that \ast indicates the corresponding author(s) but no footnote text is required. 

\noindent\textit{\small{\textbf{Received Xth XXXXXXXXXX 20XX, Accepted Xth XXXXXXXXX 20XX\newline
First published on the web Xth XXXXXXXXXX 200X}}}

\noindent \textbf{\small{DOI: 10.1039/b000000x}}
\vspace{0.6cm}
%Please do not change this text.

\noindent \normalsize{
Recent observations demonstrate that confluent tissues exhibit features of glassy dynamics, such as caging behavior and dynamical heterogeneities, although it has remained unclear how single-cell properties control this behavior. Here we develop numerical and theoretical models to calculate energy barriers to cell rearrangements, which help govern cell migration in cell monolayers. In contrast to work on sheared foams, we find that energy barrier heights are exponentially distributed and depend systematically on the cell's number of neighbors. Based on these results, we predict glassy two-time correlation functions for cell motion, with a timescale that increases rapidly as cell activity decreases. 
These correlation functions are used to construct simple random walks that  reproduce the caging behavior observed for cell trajectories in experiments.
This work provides a theoretical framework for predicting collective motion of cells in wound-healing, embryogenesis and cancer tumorogenesis.
}
\vspace{0.5cm}
 \end{@twocolumnfalse}
  ]

\section{Introduction}
%Footnotes
\footnotetext{\dag~Electronic Supplementary Information (ESI) available: [details of any supplementary information available should be included here]. See DOI: 10.1039/b000000x/}

%Please use \dag to cite the ESI in the main text of the article.
%If you article does not have ESI please remove the the \dag symbol from the title and the above footnotetext.

\footnotetext{\textit{$^{\ast}$~Department of Physics, Syracuse University, Syracuse, NY 13244, USA Fax: +1 315 443 9103; Tel: +1 315 443 3920; E-mail: mmanning@syr.edu}}
\footnotetext{\textit{$^{b}$~Syracuse Biomaterials Institute, Syracuse, NY 13244, USA }}

%additional addresses can be cited as above using the lower-case letters, c, d, e... If all authors are from the same address, no letter is required

%\footnotetext{\ddag~Additional footnotes to the title and authors can be included \emph{e.g.}\ `Present address:' or `These authors contributed equally to this work' as above using the symbols: \ddag, \textsection, and \P. Please place the appropriate symbol next to the author's name and include a \texttt{\textbackslash footnotetext} entry in the the correct place in the list.}

%%%%%%%%%%%%%%%%%%%%%
%%%%%%%%%%%%%%%%%%%%%
%%%%%%%%%%%%%%%%%%%%%
%%%%%%%%%%%%%%%%%%%%%
Many important biological processes, including embryogenesis~\cite{Farhadifar2007,Schoetz2008}, wound healing~\cite{Poujade_2007,Schneider_2010kx}, and tumorigenesis~\cite{Friedl:2009ys,irimia_toner_2009}, require cells to move through tissues.

While numerous studies have quantified cell motility by analyzing isolated cells in controlled environments~\cite{Karen_Nature_2008,Yamada_2012}, recent work has highlighted that cell motion in densely packed tissues is {\em collective}, and very different from isolated cell motion.  In densely packed or confluent tissues  (no gaps between cells) researchers have discovered signatures of collective motility such as dynamical heterogeneities~\cite{angelini_2011,KaesNJP} and caging behavior~\cite{Schoetz2013}.

These signatures also occur in many glassy non-biological materials, including polymers, granular materials, and foams ~\cite{ediger_review}. They can be understood in terms of the potential energy landscape, which specifies the total potential energy of a material as a function of the positions of all the degrees of freedom, such as the particle positions. A glassy material spends most of its time close to a mechanically stable minimum in the potential energy landscape, but rare fluctuations can overcome the high energy barriers and allow the material to escape to a new minimum.  These collective, rare fluctuations typically involve a particle escaping from a cage generated by its neighbors.

Inactive materials such as dry foams are jammed at confluence. Therefore, individual elements do not change neighbors unless a sufficient external force is applied at the boundaries.  Much effort has focused on understanding these rearrangements that occur when energy is injected globally; they tend to occur at special weak regions or soft spots in the material~\cite{manning_2011_soft_spots} and the energy barriers to rearrangements are power-law distributed~\cite{LangerLiu2000}.

Even in the absence of external forces, cells in confluent tissues regularly intercalate, or exchange neighbors~\cite{Guillot_sci_review_2013}.  They actively change their shapes and exert forces on contacts to overcome large mechanical energy barriers and transition from one metastable state to another. Because energy is injected locally, instead of globally at the boundaries, we hypothesize that the statistics of energy barriers explored by cells might be very different from those in inactive materials. The fact that glassy dynamics are observed in confluent tissues suggests that cell migration rates are governed by these energy barriers.  In other words, cell motility in tissues is set not by single-cell migration rates but instead by the rate at which cells can squeeze past neighbors. 

There is no existing theoretical framework for predicting cell migration rates in confluent tissues. Although several recent particle-based models for collective cell motion show signatures of glassy dynamics~\cite{Henkes2011, Berthier2013}, these break down at confluence and do not capture changes to cell shapes that occur during intercalation.

In this Communication, we develop a framework for predicting cell migration rates in tissues by first calculating energy barriers to cell rearrangements. We find that the distribution of energy barriers for local rearrangements is exponentially distributed, which is precisely the distribution required for  glassy dynamics in non-active matter~\cite{monthus_bouchaud}, and different from that observed in foams. Our simulation and model also predict that the height of the energy barriers depends systematically on the topology of cell neighbors in the vicinity of the rearrangement. We utilize the 'trap' model~\cite{monthus_bouchaud} and an extension of the Soft Glassy Rheology (SGR) model~\cite{SGR_1998} to convert our results for energy barrier distributions to testable predictions for cell migration, including waiting times and two-time correlation functions.  Finally, we carry out a minimal random walks based on these two-time correlation functions which capture caging and migration of cells and make qualitative comparisons to experiments.

%Shape Eq. model with homogeneous cellular properties%%
Shape equilibrium or vertex models have been successfully used to predict the minimum energy shapes of 2D cross-sections 3D cells in confluent tissues~\cite{Farhadifar2007, Hufnagel2007, Hilgenfeldt2008,manning_2010}.  These models develop an equation for the mechanical energy of a cell,
\begin{equation}
\label{total_energy}
\quad \mathcal{U}_i = \xi P_i^2 + \gamma P_i + \beta (A_i-A_{0})^2,
\end{equation}
where $P_i$ and $A_i$ are the perimeter and area of the cell.  
Coarse-grained mechanical properties of single cells that influence cell shape, which are discussed in~\cite{Farhadifar2007,ChiouShraiman}, include cortical elasticity, cortical surface tension, bulk incompressibility, and cell-cell adhesion.
The term quadratic in the perimeter accounts for the elastic contractility of the actomyosin based cortex, with modulus $\xi$. An effective `line tension' $\gamma$ couples linearly to the perimeter. $\gamma$ can be negative or positive and represents effects due to cell-cell adhesion and cortical tension. The last term quadratic in the area accounts for the bulk elasticity and additional cell-cell adhesion effects~\cite{Hufnagel2007}.
\begin{figure}[!htbp]
\begin{center}
\includegraphics[width=0.75\columnwidth]{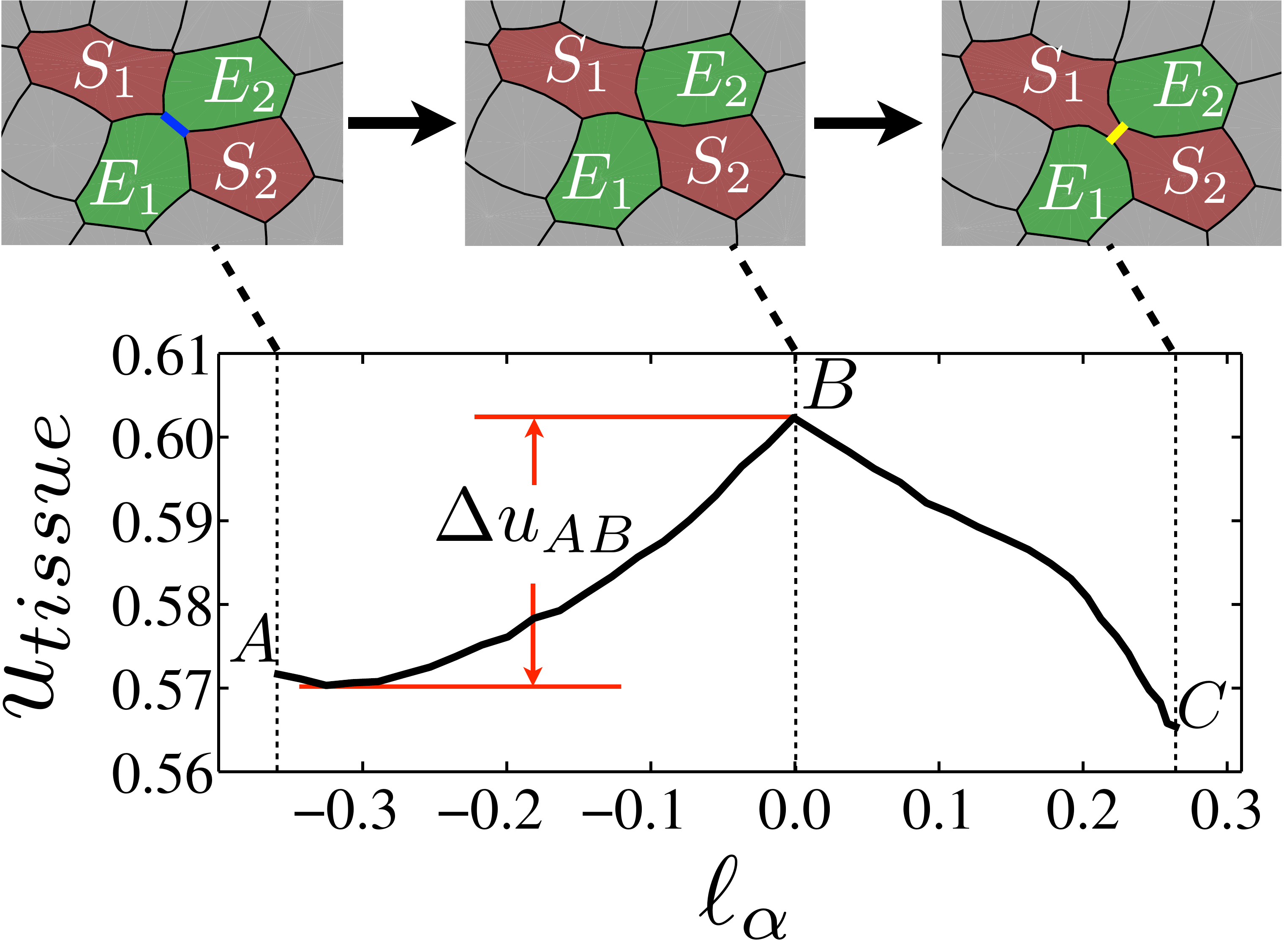}
\caption{
A T1 transition and its typical energy profile from our simulation. Cells $E1$ and $E2$ share an edge before the T1 and become disjoint after the T1, while $S1$ and $S2$ are disjoint before the T1 and share an edge after the T1. The energy increases as the edge separating cells $S_1$ and $S_2$ decreases in length, and reaches a maximum at length zero. A T1 swap takes place  and then the energy decreases as the edges separating $E_1$ and $E_2$ grows in length. The energy difference $\Delta u_{AB}$ marks the height of the energy barrier associated with this transition. 
}
\label{Figure_1}
\end{center}
\end{figure}
%%
%%%%%%%%%%%%%%%%%%%%%%%%%%%%%%
%%%%%%%%%%%%%%%%%%%%%%%%%%%%%%
%%%%%%%%%%%%%%%%%%%%%%%%%%%%%%
Quantities in Eq.~\eqref{total_energy} can be non-dimensionalized by an energy scale $\beta A_0^2$ and a length scale $\sqrt{A_0}$:
\begin{equation}
\label{total_energy_dimless}
\quad u_{tissue} = \sum_{i}u_{i}; \quad u_i= \kappa p_i^2 +2 \kappa p_0 p_i +(a_i-1)^2 ,
\end{equation}
with $\kappa = \xi / (\beta A_0)$ and $2\kappa p_0 = \gamma / (\beta A_0^{3/2})$.

This mechanical energy functional has been remarkably successful in predicting cell shapes in embryonic tissues~\cite{Farhadifar2007, Hilgenfeldt2008} and it allows for anisotropic interactions between cells.  Although a few researchers have used these models to investigate cell growth and division~\cite{Farhadifar2007, Hufnagel2007}, they have not been used to make predictions about cell migration.

Standard methods~\cite{RSA_Torquato} were used to generate a random 2D pattern of $N$ points, which was then mapped to a packing of $N$ polygons with periodic boundary conditions via voronoi tessellation. 
The program {\it Surface Evolver}~\cite{Brakke} was used to find the nearest local minimum of Eq.~(\ref{total_energy_dimless}) via a steepest descent algorithm.

Under confluent conditions, cells can only rearrange via T1 topological swaps,  as illustrated in Fig.~\ref{Figure_1}. Although cell division and death can lead to fluid-like behavior \cite{Joanny_homeostasis}, these are not necessary for cell migration~\cite{Schoetz2008, Schoetz2013} and therefore we study cell packings in the absence of these processes.
To induce a T1 transition at an edge, the total energy is minimized while the length of the edge $\ell_{\alpha}$ is actively decreased from $L_\alpha$ until the edge reaches zero length. 
Such processes are common during planar junction remodeling in epithelial layers~\cite{Guillot_sci_review_2013}.
A topological swap takes place at $\ell_\alpha= 0$. 
The new edge is actively increased to a length $L_{\alpha}$ and then allowed to relax to its final unconstrained minimized state.
Except for this T1 transition, the topology of the network of vertices and edges remains fixed.  We have also studied systems where passive energy-minimizing T1 transitions are allowed in addition to the active T1 transition, and this does not change any of the results reported below~\cite{Bi_future}. %~\cite{note_vary_topology}. 

\begin{figure}[!htb]
\begin{center}
\includegraphics[width=1\columnwidth]{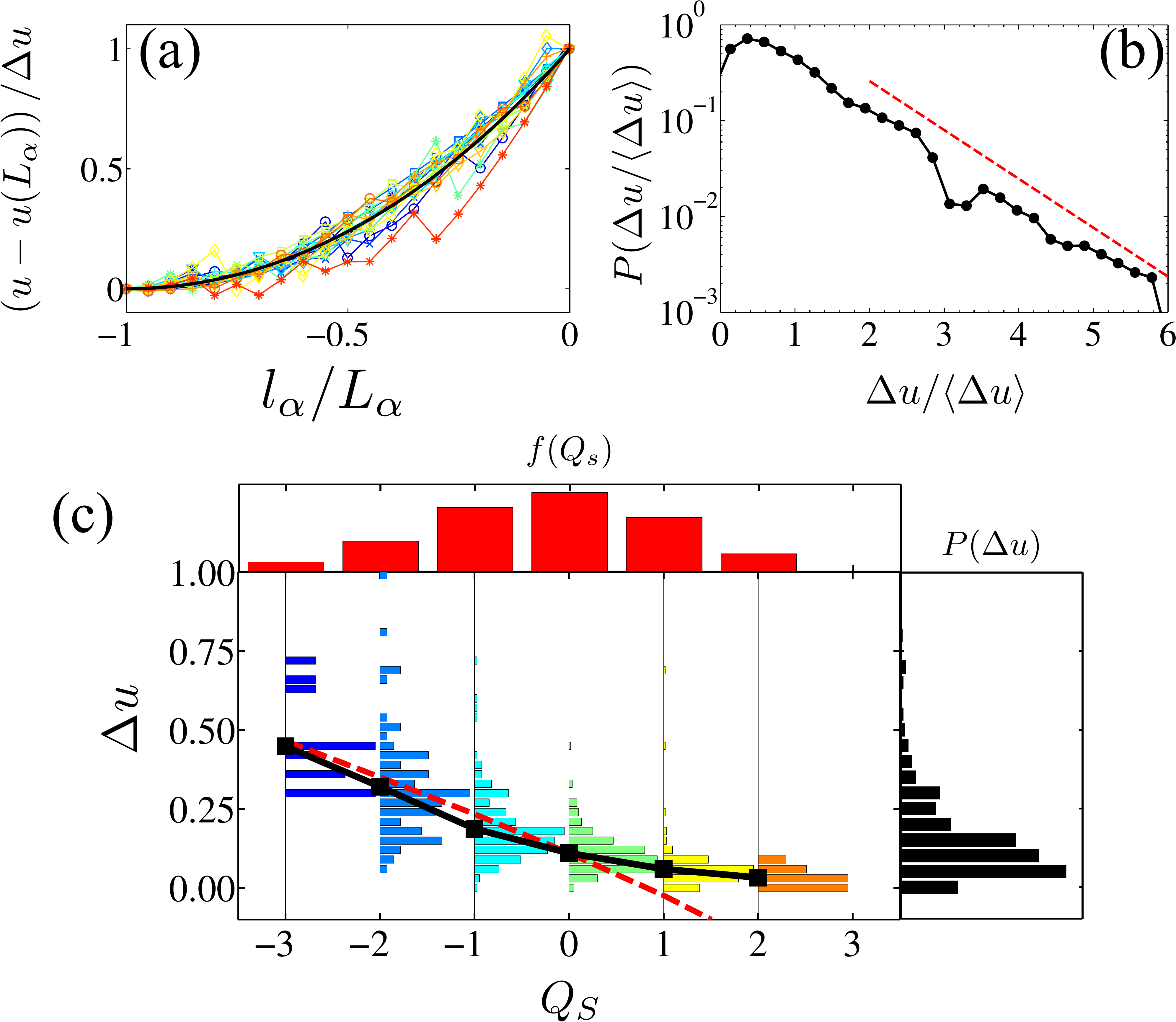}
\caption{
(a): The energy trace shows a universal behavior, as shown by the collapse of numerical results(colored thick lines) onto one curve which is predicted by the mean-field model.
(b) Probability density on a semi-log plot illustrates the exponential distribution of energy barriers. The dashed line is an exponential fit with a slope of $-1.18$.
(c)The dependence of barrier heights on the contact topology of the underlaying cells. A histogram ($p(Q_S,\Delta u)$) of energy barrier heights is shown at each value of the pre-T1 topological measure $Q_S$. Higher values of $Q_S$ correspond to $S1, S2$ pairs with fewer neighbors.
The average values are represented by the black curve. $p(Q_S,\Delta u)$ exhibits exponential tail for the range of $Q_S$ shown here. The black solid line is the average value of $\Delta u$ and the red dotted line is the meanfield theoretical prediction with no fitting parameters.  The overall distribution $P(\Delta u)$ (black histogram on right of figure) is obtained by convolving $p(Q_S,\Delta u)$ with the distribution of topological measures $f(Q_S)$ (red histogram on top).
}
\label{Figure_2}
\end{center}
\end{figure}

Fig.~\ref{Figure_1} shows the total energy of the system as a function of the edge length during a typical T1 transition. The length $\ell_{\alpha}$ is displayed as a negative number before the T1 transition and positive after the T1 transition. The energy barrier for this process  $\Delta u_{AB}$ is defined as the minimum energy required to escape state $A$ towards another stable state $C$. Statistics of $\Delta  u$ are collected by testing the T1 transition path on six randomly generated tissues each consisting of $N=64$ cells.  For all cells in a tissue, we set the parameters such that the minimal shape for each cell is a regular hexagon of area $1$: $\kappa=1$ and $p_0 \approx 3.722$. 
The distribution of energy barriers $\varrho(\Delta{u})$ of these transitions is shown in Fig.~\ref{Figure_2}(b). 
The tail obeys an exponential distribution:
\begin{equation}
\varrho(\Delta u) \propto  e^{-c \ \Delta u / \langle \Delta u \rangle} = e^{-\Delta u / \varepsilon_0},
\label{rho_dE}
\end{equation}
where fitting has determined $c=1.18$ and we define $\epsilon_0 = \langle \Delta u \rangle /c $.    
This exponential distribution is robust to changes in model parameters $\kappa$, $p_0$, cell-to-cell variations ($A_0 \to A_{0i}$) 
and the method we use to initialize cell locations~\cite{Bi_future}. 
Our data suggests that the exponential tails ultimately arise from an interplay between the statistics of edge lengths and the energy functional. Although the initial T1 edge lengths $L_{\alpha}$ are Gaussian distributed, we find that the change in energy due to a reduction in cell perimeter is quadratic in  $L_{\alpha}$, resulting in an exponential distribution for energy barriers. 

Whereas simulations of sheared foams generically generate power-law distributed energy barriers with an exponential cutoff~\cite{LangerLiu2000}, exponential energy barriers appear to be a unique feature of confluent tissues where energy is injected locally.  This is intriguing because it is precisely the distribution seen in glassy systems with quenched disorder~\cite{monthus_bouchaud}.

In~\cite{Staple2010} it was shown that the the ground state of Eq.~\eqref{total_energy_dimless} forms an ordered hexagonal lattice. However, cells in a biological tissue vary significantly in their number of neighbors or contact topologies, giving rise to a highly degenerate set of metastable states. The T1 transitions explore these metastable states and we find an interesting dependence of the energy barrier heights on the local contact topology of cells involved. As depicted in Fig.~\ref{Figure_1}, cells S1 and S2 both gain one neighbor while E1 and E2 lose one neighbor each after the transition. To quantify the dependence of the energy barrier heights on the local topology, we capture the local topology of four cells with the measure  $Q_S=(6-Z_{S1})+(6-Z_{S2})$ where $Z_{S1}$ and $Z_{S2}$ are the number of neighbors for cells S1 and S2~\footnote{The dependence on the topological measure of $E1$ and $E2$ is not included because the Aboav-Weaire law holds for our cellular packings~(Fig.~S1), and therefore the topology of $E1$ and $E2$ are strongly constrained by $Q_S$.}. Higher values of $Q_S$ correspond to $S1, S2$ pairs with fewer neighbors. After a T1 transition, $Q_S$ is always reduced by 2. In Fig.~\ref{Figure_2}(c) the energy barriers are categorized by their pre-T1 $Q_S$ values. $\Delta u$ decreases monotonically with increasing $Q_S$ and becomes vanishingly small when $Q_S=2$ (which becomes a $Q_S=0$ state after a T1 transition). This hints that the hexagonal configuration (all $Z's =6$) is not only the energetically preferred state, but configurations further away from the ground state also have higher energy barriers.

We observe that during a T1 transition most of the change in energy is localized to the four cells S1, S2, E1 and E2 that participate. Based on this observation, we develop a simple mean-field model, which considers all four cells involved in a T1 transition to be initially regular polygons of equal edge length $\ell = \sqrt{2}/ 3^{3/4} \approx 0.62$. We allow only the coordination of $S1$ and $S2$ to vary independently, and set $Z_{E1}=Z_{E2}=6$, the average value required by the Gauss-Bonnet theorem. The total energy for the four cells can be calculated for the transition path, yielding a generic profile for the energy leading up to the T1 transition,  shown by the black line in Fig.~\ref{Figure_2}(a), that is remarkably similar to simulation results. The mean-field model also predicts the energy barrier height $\Delta u_{mf}$  as a function of the topology of the cells involved, as shown by red dotted line in Fig.~\ref{Figure_2}(c).  With no fitting parameters, the mean-field model correctly predicts the magnitude of the energy barrier and the observation that lower topological measures have higher energy barriers, although it does not fit the shape of the simulation curve.  This suggests the shape of this curve is due to nontrival local correlations between cell shapes.\\

To go from energy barrier distributions to cell migration rates, 
we explore two of the simplest models to demonstrate that the observed energy barrier distribution generically yields glassy behavior, as measured by the time one has to wait to see a cell change its neighbors.  In confluent tissues, cell migration rates are then proportional to neighbor exchange rates.

In traditional statistical mechanics, the rate at which a near-equilibrium system transitions from one metastable state to another is described by an Arrhenius process, 
\begin{equation}
\label{Arrhenius}
R = \omega_0 e^{- \Delta u_{AB} / \varepsilon},
\end{equation}
where $\Delta u_{AB}$ is the energy barrier separating two metastable states $A$ and $C$ (Fig.~\ref{Figure_1})), $\omega_0$ is an inherent escape attempt frequency and $\varepsilon=k_B T$ is the scale of energy fluctuations. 

While the assumptions on which Eq.~\eqref{Arrhenius} is based do not necessarily hold in biological tissues, analogues to parameters $\omega_0$, $\Delta u_{AB}$ and $\varepsilon$ exist in cells and likely govern cell motility.  Several successful tissue models have characterized the cell activity using an effective temperature $\varepsilon$ estimated from membrane ruffling~\cite{Mombach2005}. Both $\varepsilon$ and the rate at which cells attempt to cross barriers $\omega_0$ are correlated with cell protrusivity and active shape fluctuations, which are determined in large part by the cell's individual biochemical makeup.  For simplicity, we assume that $\omega_0$ and $\varepsilon$ are single-cell properties that are constant throughout the tissue, although other choices are possible and would be interesting directions of future study.  In contrast, the distribution of energy barriers, $\varrho(\Delta u)$, is clearly a collective property determined by cell-cell interactions and the geometry of cell packing inside the tissue, as described in the previous section. 

\begin{figure}[!htb]
\begin{center}
\includegraphics[width=1\columnwidth]{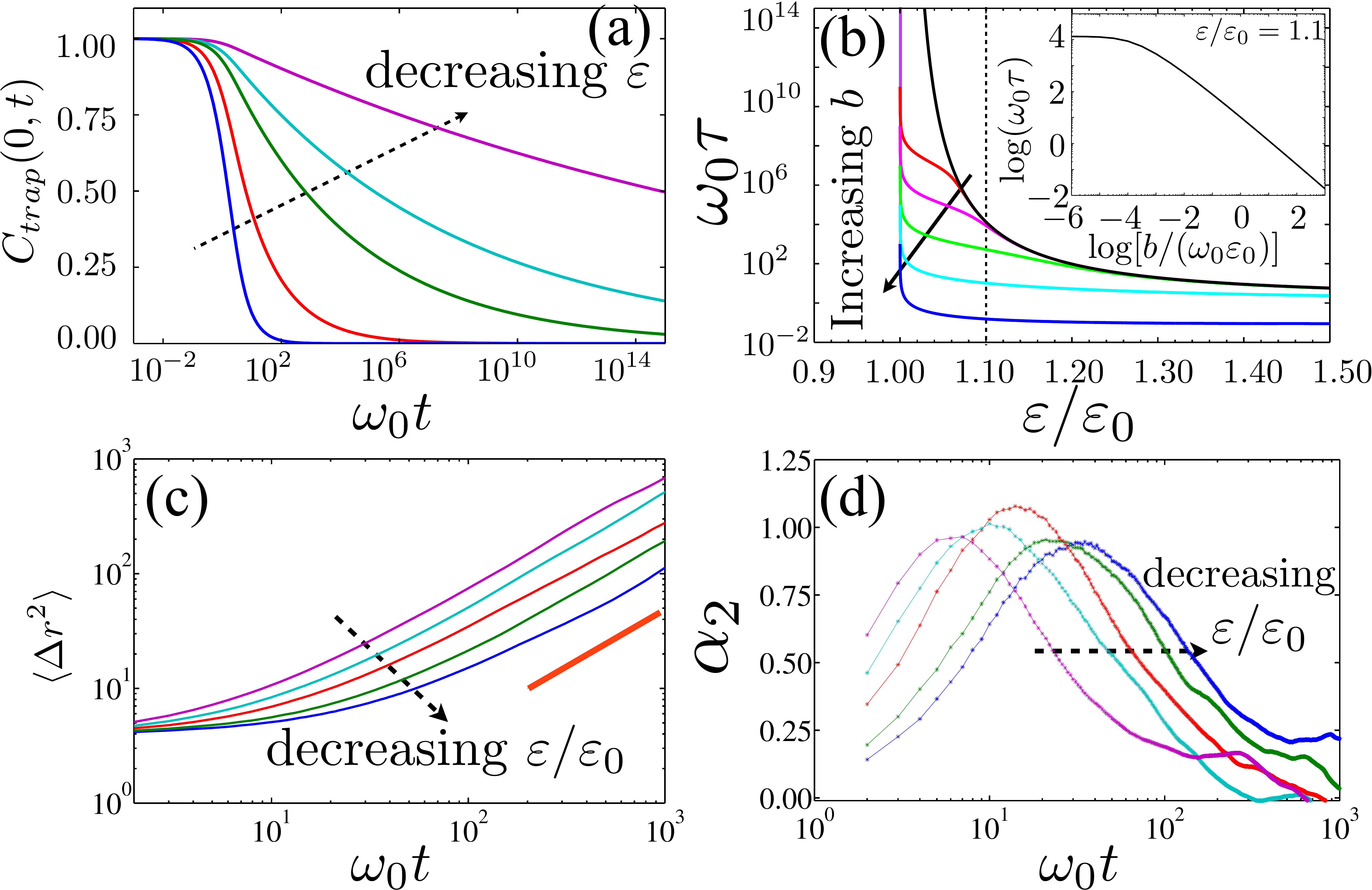}
\caption{
(a) Two-time correlation functions for $\varepsilon/\varepsilon_0 = [2.00, 1.10, 1.32, 1.06, 1.02]$  in the trap model. As $\varepsilon \to \varepsilon_0$, the correlations persist for increasingly long times, leading to glassy behavior. 
(b) Colored lines are the caging time $\tau$ in the SGR model. In the limit $b/(\omega_0 \varepsilon_0) \to 0$, the SGR model becomes the trap model (thick black line). Inset: $\tau$ as a function of  $b/(\omega_0 \varepsilon_0)$ at $\varepsilon/\varepsilon_0=1.1$ (black dashed line in the main figure).
(c) Mean squared displacement for a random walk where the step sizes are determined by the two-time correlation function $C_{trap}(0,t)$. Here we have used $b/(\omega_0 \varepsilon_0)  = 0.01$ and $\varepsilon/\varepsilon_0$ values ranging from 1.001 to 1.01. The solid red line indicates slope 1.
(d)
Non-Gaussian parameter $\alpha_2$ (described in text) for random walk tracks shown in (c). $\alpha_2$ first rises to a peak that coincides with the caging time $\tau(\varepsilon, b)$ and decays to ~ 0 as the system becomes diffusive.  $\alpha_2=0$ means diffusive behavior.
}
\label{model_glassy}
\end{center}
\end{figure}

We first use a simple `trap' model for glasses~\cite{monthus_bouchaud} to predict waiting times for cell migration.   In the trap model, a competition between $\varrho(\Delta u)$ and the Arrhenius rate (Eq.~\eqref{Arrhenius}) that samples this distribution~\cite{monthus_bouchaud} determines the dynamics. 
For tissues where $\varrho(\Delta u)$ has an exponential tail (Eq.~\eqref{rho_dE}), the distribution of the average time $\tilde{\tau}$ spent in a metastable state is given by~\cite{monthus_bouchaud}:
\begin{equation}
 f(\tilde{\tau}) \propto \tilde{\tau}^{-\varepsilon/\varepsilon_0 },
\label{waiting_time}
\end{equation}
where $\tilde{\tau}=R^{-1}$ is the inverse of the Arrhenius rate (Eq.~\eqref{Arrhenius}). When $\varepsilon < \varepsilon_0$, Eq.~\eqref{waiting_time} cannot be normalized, this means the system cannot relax to an equilibrium state, resulting in solid-like glassy behavior. 

For $\varepsilon > \varepsilon_0$, one can calculate the two-time correlation function $C_{trap}(0,t)$,  which is the probability for a cell to rearrange after spending time $t$ in a state.
In Fig.~\ref{model_glassy}(a), $C_{trap}(0,t)$ exhibits glassy or caging behavior at short times, but decays to zero at longer times, indicating fluid-like behavior. The time scale of this relaxation behavior depends on $\varepsilon$. We can define a caging time as the value of $\tau$ such that $C_{trap}(0,\tau)=e^{-1}$.
As a $\varepsilon \to \varepsilon_0$, the system approaches a glass transition and $\tau (\varepsilon)$ diverges, as shown by the black solid line in Fig.~\ref{model_glassy}(b).

We next augment this simple model to account for an additional feature of single-cell motility: single cells on substrates tend to move along the same direction for long periods of time due to polarization of the mechanical components that generate traction forces~\cite{Baker_2013, Gardel_2011}. This directed motion has been shown to be important in other models for embryonic tissues~\cite{Schoetz2013} and occurs in addition to the random fluctuations induced by changes to the cell shape that are modeled by $\varepsilon$.  Therefore we include directed cell motion in an SGR-like framework~\cite{SGR_1998}.

We use the energy barrier height $\Delta u$ to label the state of a T1 four-cell region (see Fig.~S2).  We model self-propelled, directed motion by assuming the cell by assuming that the cell actively increases the system's potential energy at a constant rate $b$. At time $t$, then the effective barrier height $\Delta u-b t$. There is also a finite probability for it to undergo a rearrangement due to non-directed fluctuations in its shape; we describe this as an activated process controlled by a temperature-like parameter $\varepsilon$~\cite{Mombach2005}. Then the rate for overcoming a barrier at time $t$ can be written as: 
\begin{equation}
R = \omega_0 e^{- (\Delta{u} - b t) / \varepsilon }.
\label{Arrhenius_mod}
\end{equation}
After escaping a trap with the rate given in Eq.~\eqref{Arrhenius_mod}, the T1 four-cell region enters into a new trap chosen from the distribution $\varrho(\Delta u)$ as given by Eq.~\ref{rho_dE}. 

Simple extensions of the SGR analysis~\cite{SGR_1998} can be used to derive $C_{trap}(0,t)$, which is again the probability for a cell to rearrange after spending time $t$ in a state. Similar to the trap model, a caging time $\tau$ can be defined.  As shown by the colored lines in Fig.~\ref{model_glassy}(b) adding a polarization energy $b$ decreases the caging time; in the limit of $b \to 0$, the SGR model becomes the trap model (a full contour plot of $\tau(\varepsilon, b)$ is also shown in Fig.~S3. In Fig.~\ref{model_glassy}(b)(inset), we show that as a function of increasing $b$ and constant $\varepsilon$, the caging time has a power-law decay.

One possible way of implementing the trap model and comparing to direct experimental results of cell motility is to carry out a  random walk using the the two-time correlation function $C_{trap}(0,t)$. First, at each time step, the state of a cell is determined by drawing a random state according to $C_{trap}(0,t)$: it is either caged with probability  $C_{trap}(0,t)$ and takes a small step chosen from a $\chi^2$ distribution or it migrates with probability $1-C_{trap}(0,t)$ and takes a larger step chosen from a Gaussian distribution. 
In Fig.~\ref{model_glassy}(c) we show the mean squared displacements of these random walk tracks near the glass transition. Cells are caged at small time scales and diffusive behavior dominates at longer times; the transition between the two regimes occurs at the time $\tau(b,\varepsilon)$ (Fig.~\ref{model_glassy}(b)). To better demonstrate cage breaking, we also analyzed the non-gaussian parameter $\alpha_2$~\cite{Schoetz2013} for these random walks as shown in Fig.~\ref{model_glassy}(d). The peaks in $\alpha_2$ also coincide with the average time of cage breaking events, directly set by $\tau(b,\varepsilon)$. As the glass transition is approached at $\varepsilon \to \varepsilon_0$, the peak shifts further to larger times, demonstrating a slowing down of dynamics in the system.
Similar mean-squared displacements and non-gaussian parameters have been seen in three-dimensional zebrafish embryos~\cite{Schoetz2013} and 2D epithelial sheets~\cite{nnetu_slow_dynamics_soft_matter}, suggesting that our simple model can explain those glassy features.

Both the trap and SGR-like models suggest that the energy barrier distribution we found in our simulations can lead glassy cell dynamics, and that waiting times for cell migration increase as the average barrier height (parameterized by $\varepsilon_0$) decreases.

%%%%%%%%%%%%%%%%%%%%%%%%%%

{\textbf{Discussion and Conclusion}}
We have simulated confluent tissue monolayers and numerically calculated the energy barriers required for cell rearrangements. We show that the distribution of energy barriers, $\varrho(\Delta u)$, is exponential and that $\Delta u$ depends on a cell's number of neighbors in a monolayer tissue. Building on these results, we show that two minimal models~\cite{SGR_1998} predict glassy dynamics, as measured by temporal correlation functions and waiting times, and a simple random walk based on these statistics reproduces features seen in experiments on confluent tissues. 

It should be possible to test these predictions in experiments on confluent monolayers.  Both the models predict that cell migration rates increase as the energy barriers decrease.  Therefore, Fig.~\ref{Figure_2}(c) predicts that cells are more likely to change neighbors if they are in regions with high topological measure (lower number of excess neighbors for S1 and S2).  Although it is difficult to track cell membranes in confluent tissues, one could estimate cell topologies by taking a voronoi tessellation of nuclei positions, and directly test this prediction. 

Furthermore, both models make predictions about  two-time correlation functions, which could be studied experimentally by looking at the decay in the overlap between a cell's initial and current voronoi areas as a function of time~\cite{abate_durian}.  One could decrease cell activity by adding drugs such as blebbistatin, and compare directly to Eq.~\eqref{Arrhenius_mod}.  
 In addition, there is a large-scale cutoff for the exponential tail in our simulations which correlates with the largest edge length in the tissue.  This suggests that in real tissues we should always expect expect the two-time correlation function to decay to zero provided one waits long enough.

Here we only model the simplest transition path leading to a T1 transition by shortening (and subsequently growing) the edges between cells. Realistically, the transition path can be more complicated. For example, protrusions can be made as the cell establishes new integrin bonds with the substrate,  developing more complicated patterns such as Rosettes~\cite{Guillot_sci_review_2013}. We have studied a few such pathways using Surface Evolver and find that they generically cost more energy, though a more systematic study is needed. In addition, we could analyze experimental cell shapes during T1 events to determine which transition pathways the cells actually take, and estimate the transition barrier across those pathways in silica.

For simplicity, our models and simulations make several assumptions about cell activity and dissipation, which should be checked and modified if necessary. For example, we assume that dissipative processes, such as the actin network being remodeled by myosin, are not strongly dependent on cell shapes/geometry and therefore we neglect them in our energy functional. This could be checked using two point microrheology, and the model could be modified accordingly.  Similarly, we have assumed that the rate at which cells attempt to cross energy barriers $\omega_0$, is also not geometry dependent. However, since mechanosensing machinery influence cell polarization~\cite{Mechanosensing} it is possible that local cell shapes systematically affect attempt frequencies, and this would be an interesting avenue of future research. Furthermore, our model postulates that the single-cell mechanical parameters $\kappa, p_0$ are independent of the activities $b$ and $\varepsilon$, but that is an assumption that we intend to relax and study.

 Finally,  in writing down trap and SGR models, we have implicitly assumed that the dynamics of cell monolayers are dominated by the potential energy landscape (like a supercooled liquid or glass), in contrast to a higher temperature normal liquid where rearrangements can happen anywhere and are not strongly constrained by the potential energy landscape.  This assumption is justified by the observations of caging behavior and dynamical heterogeneities, but also by the microscopic observation that cell structures are close to that predicted by Eq.~\ref{total_energy_dimless}~\cite{ChiouShraiman}, and transition between these near-equilibrium states quickly compared to the waiting times they spend in each state~\cite{Schoetz2013}.  Quantifying these transition times in experiments (in addition to the waiting times) would therefore be very useful.

{\bf{Acknowledgements}}
M.L.M. acknowledges the support from NSF CMMI-1334611 and the Dean of A\&S and the Chancellor's Fund at Syracuse University. J.H.L. and J.M.S. acknowledge the support from NSF-DMR-0645373.
The authors acknowledge useful discussions with Rastko Sknepnek and Shiladitya Banerjee. The authors would also like to thank the anonymous reviewers for their valuable comments and suggestions to improve the quality of the paper.


\balance


\footnotesize{
\bibliography{references} %your .bib file
\bibliographystyle{rsc} %the RSC's .bst file
}

\end{document}